\documentstyle[12pt]{article}
\input epsf

\begin{document}

% this is the def file that is called by \input{mydef}
% HOWTO notes at end of this file
% the next three defs may be bad in tex
\def \vsp {\vspace{.3cm}}

\def \vss {\vspace{.4cm}}

\def \vsl {\vspace{1cm}}

\def \fl {\flushleft}

\def \fr {\flushright}
%     particle physics definitions

\def \Snn {$\sqrt{s_{nn}}$ }
\def \S   {$\sqrt{s}$ }
\def \s   {\sqrt{s} }
\def \mM  {$\langle M \rangle$ }
\def \m   {\langle M \rangle }
\def \mpt {$\langle p_t \rangle$ }
\def \<p> {$\langle p \rangle$ }
\def \Et  {$E_t$ }
\def \pt  {$p_t$ }
\def \Pt  {p_t }
\def \et  {E_t }
\def \mex {$ E_t^\beta e^{-\alpha E_t}$}
\def \dndeta {$dn/d\eta$ }

\def \dsdn {$d\sqrt{s}/dn$ }
\def \dsdxf {$d\sigma/dx_f$ }
\def \jp {$J/\Psi$ }
\def \sjp {$\sigma_{J/\Psi}$ }
\def \up {$\Upsilon$ }
\def \Aal {$ A^\alpha (p_t) $ }
\def \Aa {$A^\alpha$ }
\def \Ae { $A^{\epsilon _x }$ }
\def \nfac {$ \frac{\langle n^2 \rangle }{ \langle n \rangle} - 1 $ }
\def \spa {$\sigma_{p-A}$ }
\def \spp {$\sigma_{p-p}$ }
\def \spn {$\sigma_{p-n}$ }
\def \Rxf {$ R(x_F) $}
\def \R   {$\sigma _{p-A} / A \sigma_{p-p} $ }
\def \Rx  {$\sigma_{p-A}(x_F) / A \sigma_{p-p}(x_F)$ }
\def \Rh  {$\sigma _{h-A} / A \sigma_{h-p} $ }
\def \Rhx  {$\sigma_{h-A}(x_F) / A \sigma_{h-p}(x_F)$ }
\def \Rpt {$\sigma_{p-A}(p_t) / A \sigma_{p-p}(p_t)$ }
\def \Rhpt {$\sigma_{h-A}(p_t) / A \sigma_{h-p}(p_t)$ }
\def \xf  {$x_F$ }
\def \fxf {$f(x_F)$ }
\def \xone {$x_1$ }
\def \xtwo  {$x_2$ }

\def \Rpi {$\sigma _{\pi -A} / A \sigma _{\pi -p} $ }
\def \dsdet {$d\sigma/dE_t$ }
\def \sjp {$\sigma_{J/\Psi}$ }
\def \dsdtau {$d\sigma/dyd\sqrt\tau = e^{-\alpha \sqrt\tau }$}

\def \pp {$pp$}
\def \pal {$p\alpha$}
\def \alal {$\alpha\alpha$}
\def \pd {$pd$}
\def \dd {$dd$}
\def \pn {$pn$}
\def \qbar {$\overline q$ }
\def \pbar {$\overline p$ }
\def \ppbar {$p-\overline p$ }
\def \qqbar {$q\overline q$ }
\def \ccbar {$c\overline c$ }
\def \bbbar {$b\overline b$ }

%statistics definitions

\def \y/ss  { \frac {y_i}{\sigma_i^2} }
\def \x/ss  { \frac {x_i}{\sigma_i^2} }
\def \xs/ss { \frac {x_i^2}{\sigma_i^2} }
\def \w/ss   { \frac {1}{\sigma_i^2} }
\def \xy/ss { \frac {x_i y_i}{\sigma_i^2} }

\def \sy/ss  {\sum_i \frac {y_i}{\sigma_i^2} }
\def \sx/ss  {\sum_i \frac {x_i}{\sigma_i^2} }
\def \sxs/ss {\sum_i \frac {x_i^2}{\sigma_i^2} }
\def \sw/ss   {\sum_i \frac {1}{\sigma_i^2} }
\def \sxy/ss {\sum_i \frac {x_i y_i}{\sigma_i^2} }
\def \chis { \sum_i{ \frac{(y_i-Y)^2}{2\sigma_i^2}} }
\def \echis {   e^{-\chis}    }
\def \qbar {\={q} }
%
%     Some average quantities (e.g., \aven = <n> )
%
\def  \avea       { \langle n_{A}  \rangle                           }
\def  \aveb       { \langle n_{B}  \rangle                           }
\def  \avec       { \langle n_{C}  \rangle                           }
\def  \aven       { \langle n      \rangle                           }
\def  \avep       {$\langle p      \rangle $                         }
\def  \aveEt      { \langle \Et    \rangle                           }
\def  \avept      { \langle \pt    \rangle                           }
\def  \dEtdphi    {  d\Et/d\phi                                      }
\def  \dEtddphi   {  d\Et/d\Delta\varphi                             }
\def  \Dofa       {  \pofa {{\partial} \over {\partial \pofa}}       }
\def  \Dofb       {  \pofb {{\partial} \over {\partial \pofb}}       }
\def  \delphi     {  \delta\varphi                                   }
%
%---------------------------------------------------------------------
\def  \etal{      {\it et al.}                                       }
%
%---------------------------------------------------------------------
%
%     Note that the following begin with 'M' ==> already in Math mode:
%---------------------------------------------------------------------

% particles
\def  \piminus    { \pi^{-}                                          }
\def  \pizero     { \pi^{0}                                          }
\def  \piplus     { \pi^{+}                                          }
\def  \pbar       { \overline p                                      }

%
%
%   Now define all (lowercase) VECTOR symbols:
%
\def  \veca       { \vec a                                           }
\def  \vecb       { \vec b                                           }
\def  \vecc       { \vec c                                           }
\def  \vecd       { \vec d                                           }
\def  \vece       { \vec e                                           }
\def  \vecf       { \vec f                                           }
\def  \vecg       { \vec g                                           }
\def  \vech       { \vec h                                           }
\def  \veci       { \vec i                                           }
\def  \vecj       { \vec j                                           }
\def  \veck       { \vec k                                           }
\def  \vecl       { \vec l                                           }
\def  \vecm       { \vec m                                           }
\def  \vecn       { \vec n                                           }
\def  \veco       { \vec o                                           }
\def  \vecp       { \vec p                                           }
\def  \vecq       { \vec q                                           }
\def  \vecr       { \vec r                                           }
\def  \vecs       { \vec s                                           }
\def  \vect       { \vec t                                           }
\def  \vecw       { \vec w                                           }
\def  \vecx       { \vec x                                           }
\def  \vecy       { \vec y                                           }
\def  \vecz       { \vec z                                           }
%
%   Now define all (uppercase) VECTOR symbols:
%
\def  \vecA       { \vec A                                           }
\def  \vecB       { \vec B                                           }
\def  \vecC       { \vec C                                           }
\def  \vecD       { \vec D                                           }
\def  \vecE       { \vec E                                           }
\def  \vecF       { \vec F                                           }
\def  \vecG       { \vec G                                           }
\def  \vecH       { \vec H                                           }
\def  \vecI       { \vec I                                           }
\def  \vecJ       { \vec J                                           }
\def  \vecK       { \vec K                                           }
\def  \vecL       { \vec L                                           }
\def  \vecM       { \vec M                                           }
\def  \vecN       { \vec N                                           }
\def  \vecO       { \vec O                                           }
\def  \vecP       { \vec P                                           }
\def  \vecQ       { \vec Q                                           }
\def  \vecR       { \vec R                                           }
\def  \vecS       { \vec S                                           }
\def  \vecT       { \vec T                                           }
\def  \vecW       { \vec W                                           }
\def  \vecX       { \vec X                                           }
\def  \vecY       { \vec Y                                           }
\def  \vecZ       { \vec Z                                           }

\def \head {\newpage
\hbox to \textwidth {{\it Chapter 2} \hfil {\it Security Issues}}
\noindent                 { \rule\.5cm] {13cm}{.05cm}}}

%
%---------------------------------------------------------------------
                            %HOW TO USE TEX

%1) type linktex
%2) make file by typing x fn tex
%3) type latex fn to compile the text making the file   fn dvi
%         tex
%4) type route typesetter to drl1200    now in profile
%5) type typeset fn dvi  to get the output.
%6) latex fn tex (only n) to print only page n

%equations    ${  }$ gets into equation mode
%             ${       }\leqno 1) $  puts eq no. on line to left.
%{\it x}  italics;   {\bf x }  boldface
%for quotes use ` which is upper left for the left hand and
%              ' which is under "    for the right hand
%\section
%\subsection

\begin{center}
{\bf  A Dependence of
Hadron Production in Inelastic Muon Scattering and Dimuon Production by
Protons}

\vsp{\bf  S. Frankel and W. Frati  }

\vsp {\bf Physics Department, University of Pennsylvania }

\vsp  {\bf Nov. 15, 1993 }

\vsl \vsl  \vsl
{\bf Abstract}

\end{center}

\vsp

         The A dependence of the production of hadrons in inelastic
muon scattering and of the production of dimuons in high $Q^2$ proton
interactions are simply related.
Feynman x distributions and z scaling distributions in nuclei are
compared with
       energy loss models. Suggestions for
new     data analyses are presented.

PACS numbers: 25.30.Mr  25.40.Ep

\newpage \centerline {\bf I. Introduction}

\vss

         When muons scatter from a proton or neutron
                        the nucleon is
excited and its
structure is changed because of the interaction of the muon with a quark
in the proton. \qqbar pairs which eventually become on-shell pions are
produced and the yield of pions can be measured as a function of z,
the ratio of the pion energy to the maximum pion energy allowable in the
process.

         The time evolution of this state just after the interaction
is unknown but it can be studied by measuring the same reaction in nuclei.
In a sense the nucleus samples the emerging hadrons because
{\it the hadrons only rarely
escape the nucleus without rescattering}. They impart
energy to the spectator nucleons and hence lose energy themselves before
coming on-shell.

    The final state  (fully clothed) hadrons
produced in this interaction are formed
roughly at the  dilated time $(\nu/m) (r_h/c)$, where $\nu$ is the
energy imparted to the nucleon         and $r_h$ is the hadron size.
  For the
data we study in this paper the energies, $\nu$, are in the hundreds of
 GeV.

Because the hadronization products cannot be calculated by perturbative
QCD we carry out
a study of {\it models} of the hadronization process
 in reactions involving
muons.

In one model the nucleon is excited and passes through the nucleus
losing energy to the spectator nucleons in its path.
 This energy loss by the excited
nucleon reduces the maximum energy available to the hadrons (mainly
pions)
it produces
and will thus affect the value of z = ${\rm E_h/E_{max}}$.
Because the muons
are weakly interacting, they only interact at their full energy
with a single target nucleon.
(The evolution is     shown in Fig. 1.)

        In a second model
     it is possible that the \qqbar pairs, which
eventually become the pions, are produced in the initial interaction.
These pairs interact with the spectator nucleons on passing through the
nucleus, imparting energy to them.
 (The evolution is shown in Fig. 2.)

        It is not difficult to estimate the effects of energy loss on
either model. What is needed is an assumption about the interaction
cross section. We can assume for simplicity that these off-shell hadrons
have a cross section, $\sigma_{inel}$,
 for reinteraction that is the measured
p-nucleon or pi-nucleon
cross section, but we shall demonstrate how the z dependence depends on
a range of   $\sigma_{inel}$'s. We also need an estimate of the energy
loss of the hadrons per collision. With this information the nuclear
z distributions can be obtained.

%\begin{figure}
%\epsfxsize=3in
%\epsfysize=3in
%centerline{\epsfbox{muhadf1.940022f}}
%caption{ }
%\end{figure}

%\begin{figure}
%\epsfxsize=3in
%\epsfysize=3in
%centerline{\epsfbox{muhadf3.940022f2}}
%\end{figure}

       In a parallel study, also involving muons, but now in the final
state, we examine the production of dimuons
by protons:  The incoming
               proton  interacts with spectator
nucleons
and   makes                   soft minimum bias collisions before making
the high $Q^2$ dimuon pair.
Thus the incoming proton's energy can be
decreased, thereby reducing the energy available for producing the rare
    dimuon pair. This
shifts events to lower Feynman x. x = ${\rm E (dimuon)/E_{max}(dimuon)}$.
(The evolution    for this reaction is shown  in Fig. 3.)

         In muon production of hadrons
the A dependence is often revealed by examining ${\rm R(z) = N_A(z)/N_D(z)
}$
while in dimuon production by protons the quantity is
${\rm R(x) = N_A(x)/N_D(x)}$. N is the number of events at either x or z.
 The comparison is usually made with the
 deuteron,
D, to include possible p-n differences.

      A most
  interesting value of R is the limiting value at z or x equal to
unity.
          In the hadron production case, if the the muon interacts with
the last nucleon in its path, there will be no energy loss. Only such
events will appear at z=1.
 In the dimuon case,
       the events at x=1 arise from those interactions where the incident
proton made the dimuon on the first nucleon in its path, the weakly
interacting dimuon pair passing unscathed through the nucleus.
If $a_n$ is the (Glauber)
probability of making n collisions in a proton-nucleon minimum bias
interaction in a nucleus, $na_n$ is the probability of making a rare
high $Q^2$ collision like a muon scatter or the production of
a dimuon pair or vector meson such as a \jp.   Since the probability of
making one collision is then $na_n/n = a_n$ the fraction of times
the collision takes place in a first or last collision is just
$ \sum a_n /\sum na_n$ =  $1/<n>$
               Thus {\it these endpoints depend only on Glauber
probabilities and not on the detailed mechanisms}. This general argument
 shows
that R cannot be a constant, independent of z or x.
 It is not difficult
to calculate these endpoint values using a Woods-Saxon spatial
distribution for the nucleons and the                        total
inelastic cross section for the nucleon-nucleon scattering.
                               That endpoint does not depend on the energy
loss function or its magnitude.

 Actually, we
really need to know the inelastic cross section for an off-shell hadron
on a ground state nucleon if we wish to calculate R at any other value
than R(1).

\vsl \centerline{\bf II. Hadron Production in Inelastic Muon Interactions}

\vsp  Figure 4 shows a plot of R(z) taken from the data of ref.
 \cite{Ryan}
The authors have chosen to separate their data sample into
a low $Q^2$ - low $x_{Bj}$ bin and a high $Q^2$ - high $x_{Bj}$ bin,
where $Q^2$ is the four momentum transfer to the nucleon and $x_{Bj}$
is Bjorken x.  As
these authors have shown, and because it can be seen that
the measured points are almost identical in the two samples,
  there is
apparently          no nuclear dependence in the data on these variables.
  As a result, and to
improve the statistical accuracy of our comparison, we have suitably
averaged the two results which are shown plotted in Figure 5.

        To examine the relative merits of fits of the data to the
conclusion in ref. \cite{Ryan}that there is no A dependence,
                 we have examined the relative $\chi^2$ for a linear
fit hypothesis to the data shown in Figure 5.
 For the fit to R = 1, the $\chi^2$ for the five measured
points is   7.2.      It is 3.0 for
the best linear
energy loss fit shown in the figure.
  (Eliminating the low statistics high z point
hardly changes the slope of the falloff but improves the $\chi^2$
by 2 rather than 1 unit.) The authors of ref. \cite{Ryan}
 remark that there are large
systematic errors in the lowest z point. Without that point
we find that the $\chi^2$ is .53 for the linear fit and 3.2 for a flat
fit with no A dependence, i.e. R = 1.
  Thus an A dependent effect seems to
provide a better
fit to the data,
          suggesting that there is a nuclear effect to be
understood.

        In several previous papers we have examined the effects of
energy loss in nuclear interactions. \cite{Dimuon} \cite{Frankel}
\cite{jaypsi}
The calculation of R(z)
is simple  if one knows the form of the energy loss per
collision. Various {\it
models} \cite{Brodsky} \cite{Quack}       have been
 proposed recently, giving different assumptions for the \S
dependence of the energy loss. We make use of
      the model
we used in 1987 \cite{Frankel} since it was demonstrated to be in
 agreement with the data observed in low \pt
production and was the
      functional form for the energy loss per
nucleon obtained by examining the energy loss found in the ISAJET
minimum bias model of F. Paige\cite{Paige},
widely used by particle physicists.
         That energy loss is approximately given by \dsdn = constant
 = $\beta$, where n is the number of collisions. Thus in our model the
energy loss varies as the square root of the laboratory energy, $\sqrt E$
  This
energy dependence lies between the values $E^0$ and $E^1$ of refs.
\cite{Brodsky} and \cite{Quack} respectively.

We first consider the model of the struck proton hadronizing outside
the nucleus into pions. In Fig. 6
we have plotted our results
for several values of ${\rm \nu = E_\mu -E_{\mu^\prime}}$.
               After these calculations were made
we obtained the actual $\nu$ distributions obtained in the experiment.
That average value is $\nu$ = 170 GeV which one can see from Figure 6
would fall well on the experimental results.
                                    All the theoretical curves predict a
rise above R = 1 at low z due to the sliding back of events to lower z
  due to energy loss.
While this appears in the data, the authors of ref. 1 caution that there
are large systematic errors in the lowest z point.
Note that                                                the point at
z = .68 falls well below R = 1.
       For this   calculation we used \dsdn = .2 GeV,
  which we had found in
 earlier
work
fit the available data on dimuon and \jp production\cite{jaypsi}.
The theoretical curves have been corrected to take into account the
bin size used in the data. The N(z) data for both Xe and D
 were fitted to a sum of two
exponentials of the form $e^{-\alpha z}$ to get the best representation of
the input to the energy loss calculation.
         To
    show the sensitivity of the calculations to both the
inelastic cross section, $\sigma_{inel}$, and \dsdn, we show in Figures 7
and 8 how variation in these parameters affects the results.
These plots also show the asymptotic limits. There is roughly a trade-off
of \dsdn of .1 GeV for a change in the inelastic cross section of 10 mbs.

%\begin{figure}
%\epsfxsize=4in
%\epsfysize=4in
%\centerline{\epsfbox{muhadf7.940022g}}
%\caption{}
%\end{figure}

         We now turn to the second calculation, namely the model in
which the \qqbar pairs are formed at the collision and lose energy on the
way out of the nucleus. Fig. 9 shows our results for an inelastic
pion-nucleon cross section of 20 mb and an energy loss \dsdn = .2
for different values of the energy
of the pion. The dependence on energy is small and all curves show a
depression below R = 1. There is a good fit to the data if we again
omit the lowest z point as suggested in ref. \cite{Ryan}.
Figure 10 shows the effect of varying the energy loss when $\nu$ = 170 GeV
and $\sigma_{tot}$ = 20 mbs.

Fig 11 shows the effect of varying the cross section of
the pion using the average value of $\nu$ of 170 GeV and the same energy
loss parameter. Here we see that 20 mb appears to give the better fit
than 10 or 31 mb. Once again we note different limits at R = 1.

          Comparing Figs 9, 10, and 11
 with Figs 5, 6 and 7, we note that in the
case of the pion energy loss the slopes of R vs z
are smaller  and the main effect is a depression of R below unity.

          We conclude that while the data are not very precise they do
not rule out the presence of energy loss mechanisms, as suggested by
W. Busza\cite{Busza}.

\vsl \centerline{\bf III. Dimuon Production in p-A Collisions }

\vsp Dimuon production in p-A collisions can be calculated on the same
energy loss model, but is slightly more complicated since the dimuon
production cross section is  energy dependent and the shape
of the cross section depends on the invariant mass of the dimuon pair.

         The effect on R(\xf) = \Rhx of any loss in energy comes about
because of the \S dependence of the Drell-Yan cross section which
varies as $e^{\gamma M/\s}$. $M$ is the invariant mass of the muon pair.
 The energy loss therefore produces two
effects a) a reduction in the dimuon yield and b) a displacement of
events to lower
$x_F$. Before demonstrating the effects of this energy
loss on R(\xf) we present a ``back of the envelope'' calculation of
R = \R ,
which shows the qualitative features.

\vsp 1) $ R = 1 - (\gamma)(M/\s)(\beta/\s)(\aven -1 )  $

\vsp This formula demonstrates how the various parameters enter into the
dimuon ``depression''.         ( Empirically the last bracket in eq. 1)
varies very roughly as ln A so, for small departures from unity, R will
vary as A raised to a small constant.)
While R should approach unity at large s, \cite{Bodwin}
it will not be unity at lab energies
as high as 800 GeV, nor will R(\xf) be independent of \xf.

      We now turn to two pieces of data that illustrate the effect of
energy loss. We use the Drell-Yan formalism and the Duke-Owens form
 factors for the calculation\cite{Duke}, rather than using the empirical
values for the energy dependence of the cross section, since we have
 checked that they give essentially the same results.

Fig. 12
 shows the data of the NA10 collaboration on dimuon production by
140 and 286 GeV pions\cite{Bordalo}.    Superimposed on
their data are our calculations for .2 and .4 GeV energy loss. There
appears to be a clear M dependence             as well as deviations
from R=1 in the \xf distributions.  The data is not very precise but the
general features of energy loss are borne out.

%\begin{figure}
%\epsfxsize=4in
%\epsfysize=4in
%\epsfbox{muhadf12.0065b}
%\end{figure}

Fig. 13 shows the 800 GeV data\cite{Alde}
of Fermilab expt. E 772. We have obtained the unpublished mass
distributions \cite{Peng}                                  for each \xf
bin so we can make a comparison between the data for different values of
\dsdn. The published \xf distributions for their W to Deuterium ratios
  show a deviation in R from unity at large \xf. Our calculations
for \dsdn = 0.2 are superimposed on the data.

%\begin{figure}
%\epsfxsize=4in
%\epsfysize=4in
%\epsfbox{muhadf13.0086a,b}
%\end{figure}

         We conclude that there are A dependent effects in dimuon
 production
which can be accounted for using the same energy loss
mechanism that can account for
the A dependence in inelastic muon scattering  and using the same
approximate
 energy
 loss parameter.

\centerline{\bf IV. Discussion}

\vss
         Unfortunately, none of the data extend to very large values of
z or x to enable the asymptotic values at R(1) to be compared with the
Glauber prediction, verifying the importance of the energy loss in the
 most
unambiguous way.  There are other recent
estimates for the functional form of
the energy loss \cite{Brodsky} \cite{Quack} but the present data cannot
easily be used to discriminate among the various models.

  However the question of the form of the energy dependence of \dsdn
can be more
easily examined with present data on hadron production by muons.
   E665 can use all its data not just
their published low and high $Q^2$ - Bjorken x bins to improve the
statistical accuracy for a new study:
                              the
             separation of
 the present E 665 muon  inelastic data into
separate $\nu$
bins to  study R($\nu$). Newer E665 data could be analyzed to study R
 for a wide range of A. For example, with our
parameterization \dsdn, the laboratory
           energy loss, dE/dn, would vary as $\sqrt
 E$
so that the loss parameter would clearly change in the region covered by
 the
E665 data (110 to 490 GeV). While it is difficult to study the energy
dependence of \dsdn in dimuon production, since different accelerator
energies are needed, it is simple to select $\nu$ from the inelastic
muon data to make the same study.

         This work extends to muon inelastic scattering  and dimuon
production
          the need for energy loss in traversing nuclear matter
in accounting for A dependences which we showed earlier \cite{Frankel}
could account for various A dependences
        in minimum
bias reactions.

\vss  \centerline{\bf  V. Acknowledgements }

\vsp  We should like to thank J. Ryan and W. Busza for useful
and spirited discussions.

\newpage \centerline{\bf Figures }

\vss
Fig. 1 Hadron Production by Muons: An incident muon interacts with a
nucleon which rescatters off another  nucleon in the nucleus. The
nucleon struck by the muon hadronizes outside the nucleus.

\vss
Fig. 2 Pion Production by Muons: An incident muon interacts with a
nucleon producing an off-shell pion which rescatters off nucleons in
the nucleus

\vsp
Fig. 3 Dimuon Production by Hadrons: An incident proton makes soft
non-perturbative scatters before making a high $Q^2$ collision with
another nucleon producing an excited proton which hadronizes outside the
nucleus.

\vsp
Fig. 4  The z dependence of the ratio of hadron production in Xenon and
Deuterium,  R(${Xe}/{D}$),
 for high (open circles)
and low (solid circles)
$Q^2$ and $x_{Bj}$.
                               The data are  taken from ref. 1.
Table XXXI.

\vsp
Fig. 5  R vs z for the combined sample of high and low $Q^2$ data.
The solid curve is a least-square fit to a straight line fit to the data.

\vsp
Fig. 6 Hadron Energy Loss:
 R vs z for the combined sample of high and low $Q^2$ data.
Theoretical curves are shown for three values of the energy transfer,
$\nu$. The mean value of $\nu$ for the data is 170 GeV.

\vsp
Fig. 7 Hadron Energy Loss:
 7 Theoretical calculations of R for different values of \dsdn.
The inelastic p-p cross section is set at 31 mb. The value of R at
z=1 is set by this cross section and the Woods-Saxon nucleon spatial
distribution.

\vsp   Fig. 8
Hadron Energy Loss:
 Calculations of R for different inelastic cross sections. The
value of \dsdn = .2 GeV. Note the different asymptotes.

\vsp
Fig. 9 ``Pion Energy Loss'':
  R vs z for the combined sample of high and low $Q^2$ data.
Theoretical curves are shown for three values of the energy transfer,
$\nu$. The pion-nucleon cross section is taken as 20 mb. \dsdn = .2.
The mean value of $\nu$ for the data is 170 GeV.

\vsp
Fig. 10 ``Pion'' Energy Loss:
Theoretical calculations of R for different values of \dsdn = .1, .2,
.3, and .4.
The inelastic pi-p cross section is set at 20 mb. The value of R at
z=1 is set by this cross section and the Woods-Saxon nucleon spatial
distribution.

\vsp   Fig. 11
``Pion'' Energy Loss:
 Calculations of R for different inelastic cross sections, 10, 20 and 31
mb.
  The
value of \dsdn = .2 GeV. $\nu$ = 170 GeV.
 Note the different asymptotes.

\vsp

Fig. 12 Data of ref. \cite{Bordalo} for dimuon production by pions.
The predicted falloff of R with the dimuon invariant mass, M, as well
as with Feynman X are shown along with the theoretical energy loss
predictions.

\vsp
Fig. 13 Dimuon production by 800 GeV protons. R vs \xf. Data
 from ref.\cite{Alde}

\flushright \small  muhad4ds.tex 8-94
\end{document}